# Role of Project Management in Virtual Team's Success

**Attique Ur Rehman**
Department of Computer and
Software Engineering,
CEME,
NUST
Islamabad, Pakistan
aurehman.cse19ceme@ce.ceme.edu.pk

**Ali Nawaz**
Department of Computer and
Software Engineering,
CEME,
NUST
Islamabad, Pakistan
anawaz.cse19ceme@ce.ceme.edu.pk

**Muhammad Abbas**
Department of Computer and
Software Engineering,
CEME,
NUST
Islamabad, Pakistan
m.abbas@ceme.nust.edu.pk

**ABSTRACT**
A virtual team is a group of geographically distant people who work together to achieve a shared goal for a common organization. From the past few years this concept has been evolved and has emerged the idea of global project management. Virtual teams have been beneficial in cost reduction; hiring competent work force and improving globalization. Although virtual teams are beneficial for an organization; but they are hard to manage and control successfully. There can be several challenges like cultural issues; different time zones and communication gap. These challenges are not hard to manage; and we can overcome these challenges using effective project management skills. These skills will become the success factors for making virtual teams successful and will be determined by comparison of the survey results of traditional and virtual teams.

## 1. INTRODUCTION

The world has become a global village through advanced communication. This has not only facilitated for more connections and communications around the world, but people have used this advancement for other processes. So much so that people prefer to work with other people around the world despite of geographical distances.

Virtual teams are made for this purpose. The virtual team or VT is teams which share responsibilities and strive for a common goal while physical meetup is not possible [1]. These teams are formed by adding people in a group who often find themselves different in terms of time, culture and they are geographically distant, yet perform to achieve a common goal [2]. There can be different cases of being geographically distant. In some cases only the leader can be in a certain place and the team members can be in some other place while in some cases the leader and some of team members might be in one place where as other people will be geographically distant or there can be a situation when a leader with more number of team members will be at one place and the rest of the teams i.e. less in number can be at different number of geographically distant areas [3].

Virtual teams have become an important part of many large and small firms worldwide. A lot of people are working in remote teams to achieve certain objectives. This concept of virtual team has not only evolved communication, but it has also enhanced much more business opportunities for people around the world. Having virtual teams brings exposure to your business from information and cultural point of view. Other than these points it has helped a lot of companies and firms in many different aspects some of which are mentioned below.

- Office Cost Reduction
- Enhanced productivity
- 24/7 working possibilities
- Exposure to talent
- "Work from Home" concept
- Lower salaries
- Reduced meetings
- Lower travelling costs

Although virtual teams are enhancing the business profits and increasing the communication and exposure around the world but there are certain scenarios in which virtual teams have failed badly. The reason behind their failure can be poor project management and poor project managers. Sherri Mackey, The Global Coach who is founder of Luminosity Global Consulting Group, Global Executive Coach, Speaker, Writer and Global Business and Cultural Expert, states in her blog that the total failure rate of virtual teams may be as high as up to 70%. Some of the issues of Virtual Teams are described further [4].

## 2. Literature Review

Global software development (GSD) is a software development paradigm in which development activities are carried out by experts who are based in different locations around the world in order to develop successful products for a corporation [8]. GSD causes a few issues for experts that don't exist in programming ventures that are created at the collocated place [9]. Group is situated in a few diverse physical areas, contrasts in ethnicity and time zones adverse effect correspondence and coordination [10]. In software Industry Worldwide virtual groups are getting increasingly basic as associations contend comprehensively [11]. As rapid increment in global virtual teams the major issue arise are basically due to two main factors

- Lack of communication
- Lack of collaboration

Lack of communication always leads to collaboration problems [12].

### 2.1 EXTRACTED ISSUES:
Working in a team and making communicating more effective remains always challenging, but when we talk about remotely located teams these issues always require attention and remains in concern [13]. As we know that regular communication of people in organization is necessary and important but now a days in most organizations it becomes more difficult to keep in touch with team members on regular basis. One trend that is outsourcing is becoming more common today; members or concerned domain consultants are added as a member of project team, who are not a regular employees of parent organization [14]. They get hired and becomes team members as they are skilled or have expertise that is essential for the project team but currently not available or on a tight schedule inside the organization. Another trend is involving workers in many project teams and at same time they are working parallel on different software projects simultaneously. It has become common for projects that large, dynamic in nature, required evolution over time and change over the development life cycle of the project. Today most of the teams are facing these difficulties, in addition to these problems we have the concept of distributed teams, working from different areas of same country, different countries with their respective time zones, having different cultural or business languages. We can see how the continuously levels of difficulty is being increased and creates issues and challenges in managing virtual or remote teams. Some common challenges that virtual teams have faced are gathered by researchers1 are described here.

- Building trust within virtual teams is difficult due to lack of interaction.
- Virtual team members feel isolated and detached because they are not physically involved in social activities of the organization.
- Establishment of good group dynamics and it is sustainability becoming more difficult and complex in any virtual team
- Establishing consistent roles and expectations is challenging, virtual teams will have to struggle for creating synergy.

Virtual communications cannot be a replacement of physical communications. No one can deny the importance that possess face to face meetings and no advanced communication technology can replace it.

- Assessment and reporting are limited because managers need to develop enhanced approaches so that feedback and support may provide to project's virtual team.

We have discussed some common issues, these issues don't make the development process less or non-important, but the process becomes more complicated. And these highlighted issues must be dealt with serious concerns in order to make virtual team more successful and efficient. So, here are some success factors on which mangers can count while dealing with virtual teams.

## 3. SUCCESS FACTORS

As we mentioned earlier, there can be certain issues for making virtual teams more successful, but these are not completely responsible for failing the process completely. So, we have defined certain success factors in the light of the knowledge areas of Project Management. We have categorized the success factors under the knowledge areas to be more specific about them. This type of categorization is always helpful in implementing those success factors. Therefore, following are the success factors under the categorization of the knowledge areas of Project Management.

### 3.1 Communication Management:

This is the most important knowledge area in reference to the success for Virtual Teams. As we know that communication is the most crucial activity/aspect/criteria for the success of any project thus we can differentiate it into two categories in accordance with the types of teams a project manager can have.

- Communication with Co-located Teams
- Communication with Virtual Teams

This is obvious that although communication with co-located teams requires good communication skills and some advanced techniques to handle the team, yet it also provides more chances of face-to-face meetings which is always helpful in getting a better response from the team. It also brings more collaboration from the team as well. Whereas in case of virtual teams, communication is more technology dependent. All the communication processes are performed and accomplished with the help of technological means which obviously don't show much of a participation of the team and the project manager is also void of the idea related to the responses of the teams. The following diagram gives a picture of difference of communication between both types of teams. So, to communicate better with virtual teams, we shall try first to remove the differences as much as possible. Other points are also mentioned which can help us overcome these differences and can help us communicate better with the Virtual Teams.

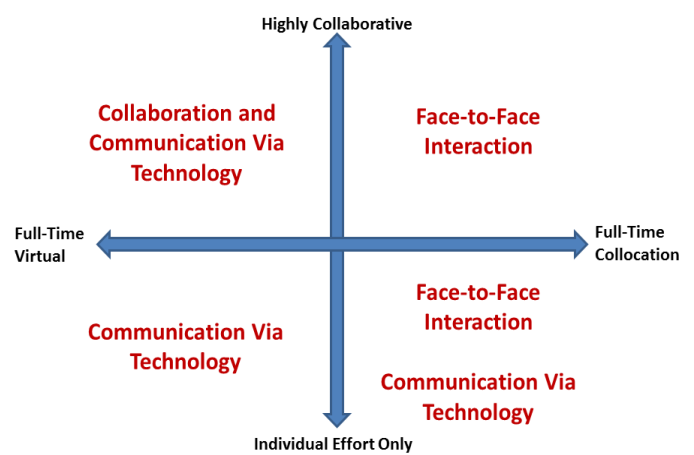

**Figure 1: Communication and collaboration matrix [6]**

### 3.2 Keeping the Team Engaged

One of the most important tasks is to keep the team engaged with the project, the company and their tasks for the success of the project. The teams shall be often updated regularly for the success of project being developed. This is the task of the project manager to keep the team updated and keep them involved and engaged in their tasks and in their tasks so that this could enhance their performance in more effective way. A project manager shall also keep in mind that each member is important, so it means that he is supposed to communicate with all the members of the virtual teams. This will bring an enhanced performance out of the people.

---

[1] Challenges in managing virtual teams

Rudy Nydegger, Ph.D., Union Graduate College, USA

Liesl Nydegger, B.A., Claremont Graduate University, USA

Journal of Business & Economics Research – March, 2010

## 3.3 Improved meeting criteria and methods:

Meeting are very important part of communication. Proper meeting standards shall be developed in order to achieve positive meeting outcomes Meeting is something that has been always an important issue for the communication between virtual teams.

This has been such an issue because in most of the case, no proper standards have been maintained to have proper meetings. In most of the cases a simple video call connecting a project manager and any member of the team is considered to be the meeting. Clearly, these are not quite successful. There can be the following points which can achieve a better standard for meetings.

- Set proper schedules for meetings
- Set rules for meetings as per the policy of the organization or as per the requirement of the project manager
- Conduct face-to-face meeting sessions once in a month, if possible

## 3.4 Build Strong Trust:

Lacking trust is a strong issue that causes a failure for the virtual teams. Therefore, there can be certain targets which can be achieved to develop trust amongst the virtual team members, as we know that co-located teams don't face this situation as severely as the virtual teams do. The team leaders should make sure of the following points [5].

- Provide opportunities for social interaction
- Break the Ice amongst members via different exercises
- Provide certainty about their individual roles

## 3.5 HR Management:

HR management is also very crucial activity when it comes to managing the virtual teams. The project manager should must make it sure that he is managing his human resource efficiently in order to get successful virtual teams. The following factors shall be considered in order to have successful process of managing virtual human resource.

### 3.5.1 Hiring Team Members

Hiring team for a project is obviously a very crucial task and the project manager requires a good experience and sound knowledge for this purpose. The project manager must keep in mind the following point if he wants to have a successful team.

- Hire the right person for the right job
- Hire trustworthy people
- Initially the hiring should be a trial
- Hire self-motivated people
- Hire the people with strong technological background

### 3.5.2 Encouraging Policies and Strong Career Opportunities

Project managers should try to make policies for the virtual teams which could motivate them to perform more efficiently. Similarly, there shall be more opportunities for the virtual team members those shall encourage them to stick to the team and not look for other jobs. The following points shall be considered.

- Provide promotions for the virtual team members
- Provide career options
- Encourage employees by identifying their significance towards the project the organization
- Provide benefits and funds to virtual team members
- Provide special trainings similar to those of co-located teams
- Provide non-bureaucratic environment i.e. provide flexibility and freedom

## 3.6 Procurement Management

In most of the cases there can be certain scenarios where there is important manage your resources efficiently so that your team can have the most of those resources. Consider the following points in order to have efficient utilization of your resources.

### 3.6.1 Technological Resources

As a project manager you should make sure that you provide the technological resources to your team.

- Provide collaborative tools for better collaboration e.g. Asana
- Provide tools for better meeting sessions and video conferences
- Provide tools which could enhance the developmental skills of the members
- Provide tools for efficient accomplishment of other processes of the project

### 3.6.2 Workspace Resources

In most of the cases, virtual teams do work from home. But this often leads to a sort of de motivated situation as a person requires a proper working environment to give a productive outcome. So, the project manager should make sure of the following points.

- Provide a proper workspace/office for the team
- Provide necessaries wanted by the employees e.g. WIFI, electricity etc.
- Provide recreational resources in the workspace

## 3.7 Scope Management

Scope management can be very important for being the part of success of the virtual teams. The leaders of the team should keep the scope in accordance with the following points.

### 3.7.1 Narrowed Scope and Clearly Defined Objective

The project managers should keep the scope narrowed for the project. It will be good for the project over all as well as the success of the virtual teams. Keep in mind the following points.

- Eliminate the un-necessary details for better understanding
- Define the objective clearly for performing to the point tasks

These are the success factors defined in the light of the project management knowledge areas. These success factors have been validated by performing surveys and calculating results from those surveys which are described in the next part of the paper.

## 4. ANALYSIS

The main purpose and goal of this research is to explore the challenges of software project management in a virtual or remote working environment, to focus on how the highlighted challenges regarding to a virtual context affect relationships and performance of virtual team, how to build mutual trust within members of virtual teams and how higher leadership influence a virtual settings. Initially through theoretical framework we have concluded some issues regarding virtual teams that helped us in making questionnaire. We distributed our questionnaire to different People who worked remotely for some organizations. Next, we studied each questionnaire to conclude the success factors for managing virtual teams. Research related details are given below.

## 4.1 Background

Questionnaire was prepared through literature review related to virtual team management issues. Then we selected individuals to solve this questionnaire who worked remotely. The 25 respondents in our survey were based in 12 different companies working remotely as presented in Table 1. Also, their roles among the organizations are mentioned in Table 2.

**Table 1: participating companies**

| No | Company Name | Individual |
|---|---|---|
| 1 | Wanclouds Inc | 11 |
| 2 | Bullseye communication | 1 |
| 3 | Web Evangelist / Freelancer | 1 |
| 4 | Access 360 | 1 |
| 5 | Red Buffer | 2 |
| 6 | Sherserve | 1 |
| 7 | Tristars | 1 |
| 8 | ZSoft | 1 |
| 9 | Nokia Networks | 1 |
| 10 | Freelance Graphic Designer | 1 |
| 11 | ZSoft | 1 |
| 12 | ZeroPoint | 1 |

In addition, concerning the amount of time respondents worked remotely in virtual projects (Figure 1), 32% of the respondents indicated less than 40 percent, followed by 20% of the respondents indicating 40-60 and 60 – 80 percent of their time, 16% of them indicated at least 80 percent of their time and only 12% reported virtual projects as the only way they worked.

**Table 2: Employee roles**

| No. | Functional role | %age |
|---|---|---|
| 1 | General management | 16% |
| 2 | Human resource | 8% |
| 3 | Development team | 8% |
| 4 | Quality assurance | 4% |
| 5 | Finance | 2% |
| 6 | Risk | 0% |
| 7 | IT/Communication | 40% |
| 8 | Strategy and business development | 12% |
| 9 | Procurement | 0% |
| 10 | Others | 8% |

The mentioned aspect is also a key important and significant factor in this research, as it emphasizes the major fact that respondents were involved both in physically present and virtual projects, it will lead us to have a better understanding regarding the challenges and key success factors of working in a virtual project team environment. This will also show the importance of partial face-to-face communication.

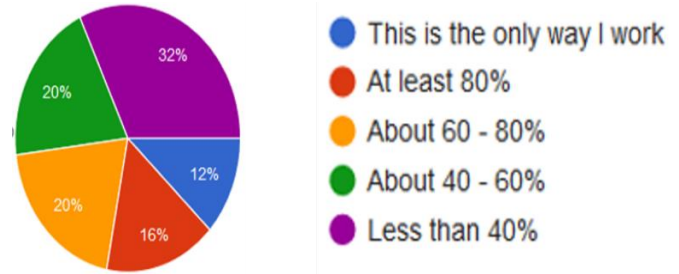

**Figure 2: Survey Response**

## 4.2 The challenges enabled by virtual teams

According to many team respondents the most challenging and significant factor for virtual or remote teams was major difference in time zones (72%). Coordination of tasks (32%). Slightly less but not least challenging were the cultural differences (28%) and trust building (24%) and the factors considered the least challenging were computer-mediated communication (16%) and language (12%). But language proficiency difference affects project's individual effectiveness and enable communication barriers [7]. Answering these questions mostly respondents were neutral. Another category of questions was how the respondents find their team leads expertise and his responses to their queries. Level of disagreement was more in this concern. Finally, we inquired about the respondent's satisfaction about the team, project and the team lead. For further detailed answers are provided in Questionnaire and response section.

For mean value analysis of the results obtained we assigned some weights to the options provided against each question. Options with Further on in the questionnaire that was based on agreeing or not agreeing about certain conditions. First two questions were trust related 56% and 48% of the respondents agreed about clear role definition and trusting other members of team and their work respectively. While 12% and 8% experienced trust issues and disagreed or strongly disagreed, 16% and 24% were neutral. These findings tell us that the most respondents have no trust related issues in their virtual teams. Next four questions were about communication and collaborative tools used by the team lead. Answering these questions mostly respondents were neutral. Another category of questions was how the respondents find their team leads expertise and his responses to their queries. Level of disagreement was more in this concern. Finally, we inquired about the respondent's satisfaction about the team, project and the team lead. For further detailed answers are provided in Questionnaire and response section.

For mean value analysis of the results obtained we assigned some weights to the options provided against each question. Options with their assigned weights are given below

- Strongly Disagree response will be ranked 1
- Disagree response will be ranked as 2
- Neutral response will be considered 3

- Agree response will be ranked as 4
- Strongly Agree will be 5

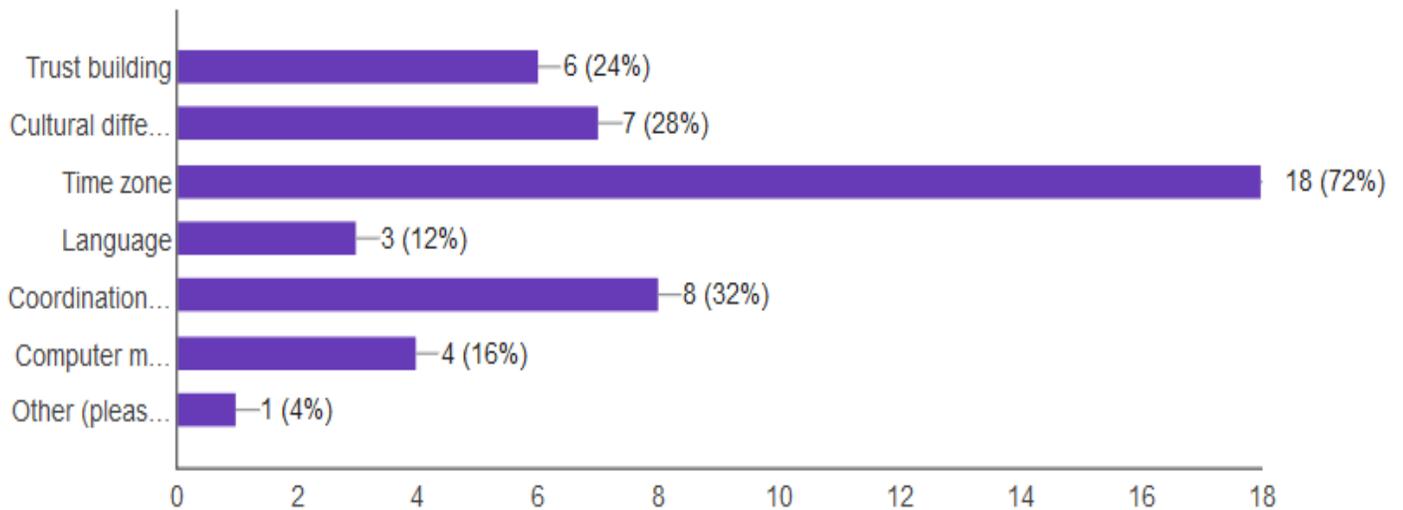

Figure 3: Challenges response graph

## 5. QUESTIONNAIRE RESPONSE

### 5.1 A role I had is clear and assigned tasks are also clear within team.

*Mean: 3.68*

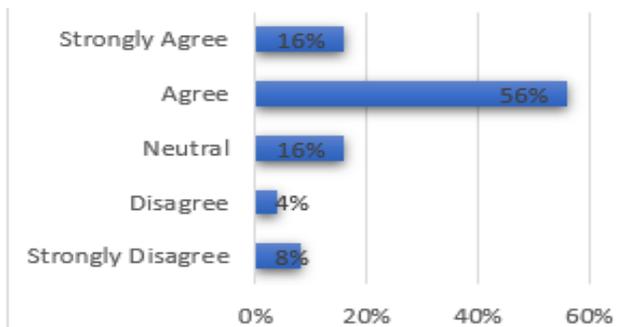

Figure 4: Response graph 5.1

### 5.2 I receive prompt response on my query by my team leader

*Mean: 3.76*

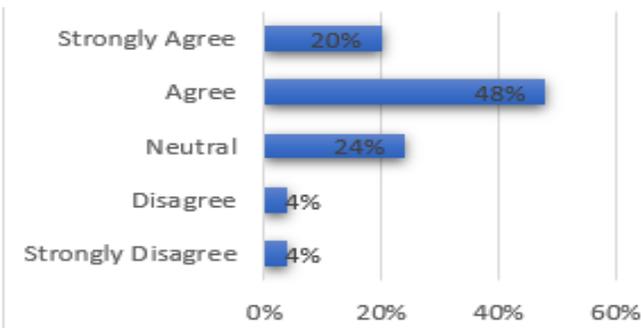

Figure 5: Response graph 5.2

### 5.3 An unbiased with high accuracy of feedback that I received from my team leader is about individual person's and team performance been implemented by my team leader within team.

*Mean: 4.08*

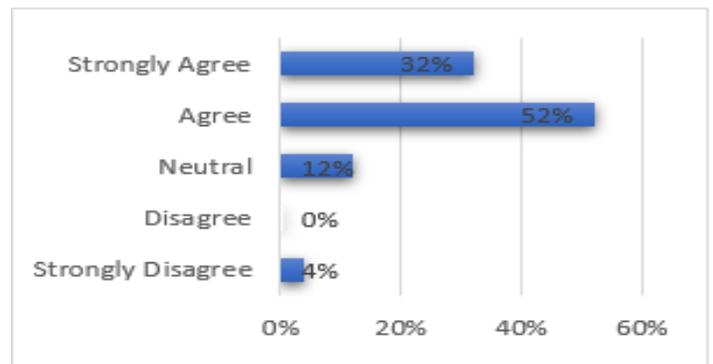

Figure 6: response graph 5.3

### 5.4 A communication guideline has been implemented by my team leader within team.

*Mean: 3.92*

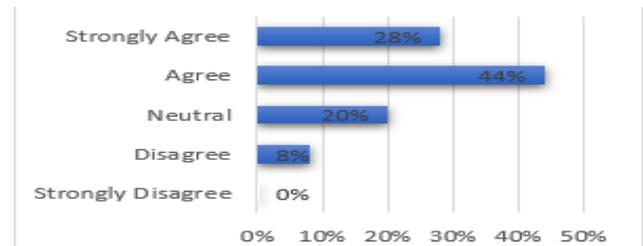

Figure 7: Response graph 5.4

**5.5 The Collaboration tools that are advanced and effective are being used by my team leader.**
*Mean : 3.76*

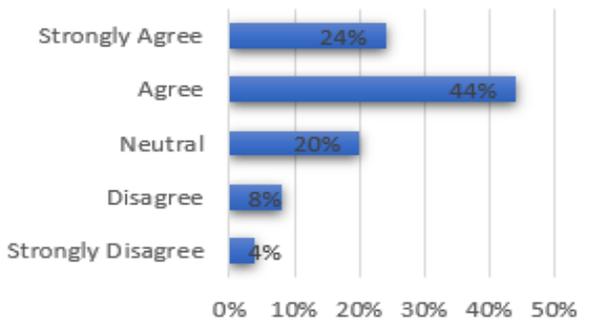

Figure 8: Response graph 5.5

**5.6 For help and support I felt other team members could be trusted.**
*Mean : 3.40*

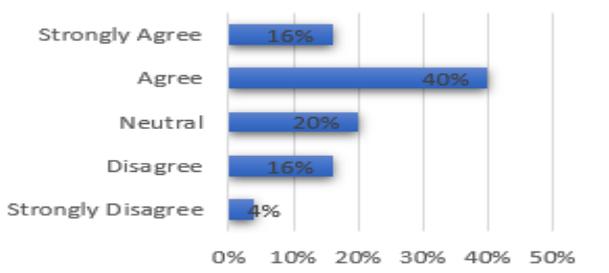

Figure 9: Response graph 5.6

**5.7 Other than official work exchange of information is frequent in my team**
*Mean: 3.4*

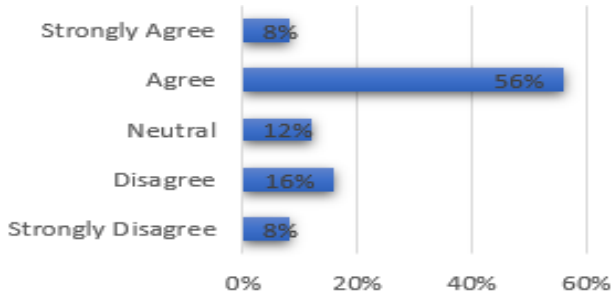

Figure 10: Response Graph 5.7

**5.8 Misunderstandings or other project conflicts are often faced by my team.**
*Mean: 3.36*

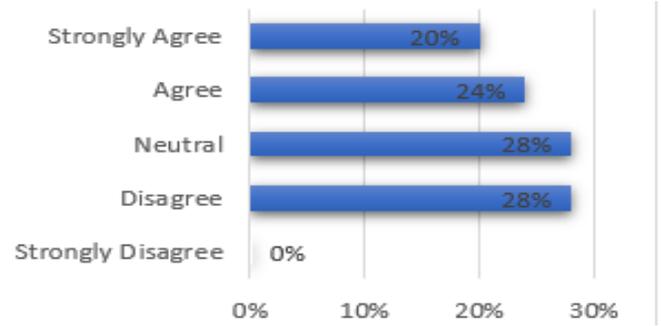

Figure 11: Response Graph 5.8

**5.9 My individual achievements with respect to other team members are acknowledged by my team leader**
*Mean: 3.72*

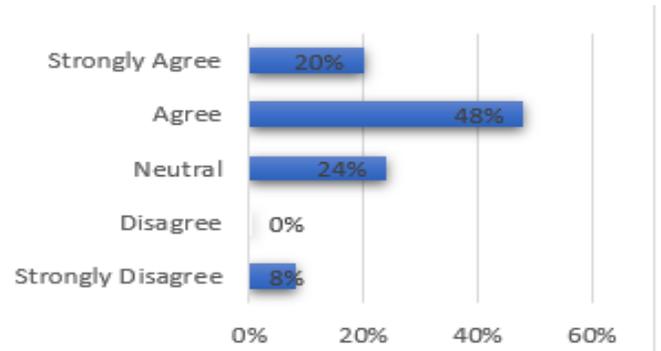

Figure 12: Response Graph 5.9

**5.10 Project Goals are being achieved with in schedule by my team**
*Mean: 3.4*

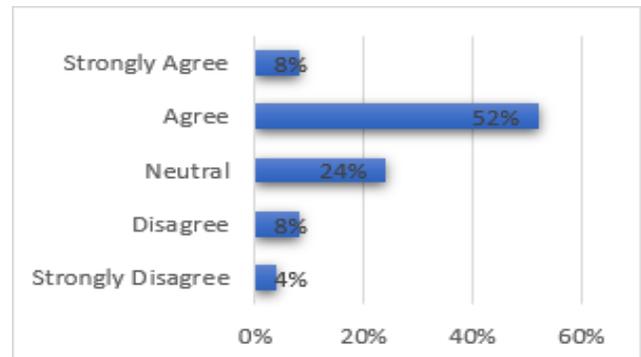

Figure 13: Response Graph 5.10

## 5.11 The way of leading adopted by project manager for virtual team setting is suitable for project.

*Mean: 3.48*

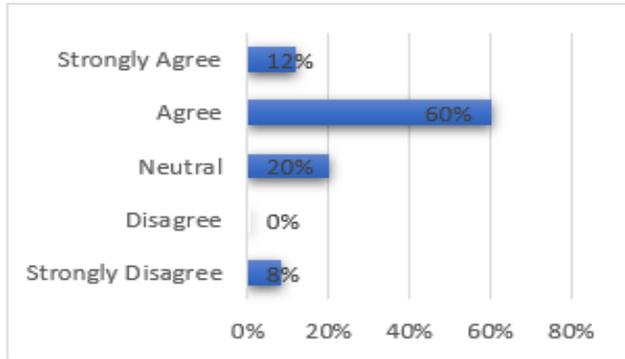

**Figure 14: Response Graph 5.11**

## 5.12 It is felt by me that all my other team members in virtual team is well known to me?

*Mean: 3.40*

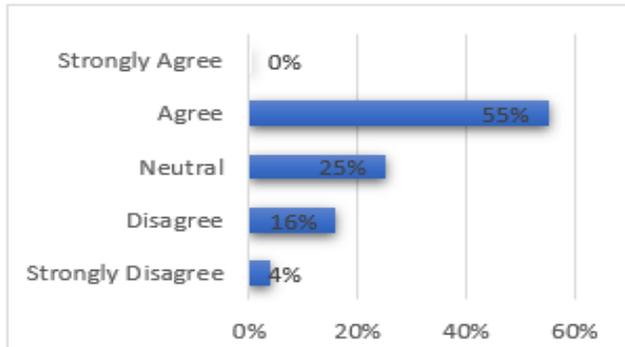

**Figure 15: Response Graph 5.12**

## 6. CONCLUSION AND FUTURE WORK

This research highlights the issues faced by project development organization in the context of GSD. Organizations. Members of virtual team need to have effective communication and collaboration in order to make the project activities successful. the study helps to identify major communication flaws and their effects on collaboration this study identified that the effective collaboration is always dependent on effective communication. Lack in communication and collaboration affects all knowledge area of software project management that leads to failure of the project. There is need of some counter measures that should be undertaken to make communication and collaboration effective although there is a complete process of communication management but still there is a gap of collaboration management there is need of collaboration management in order to make software development more effective in the context of GSD. But collaboration management could be applied categorically on two kinds of projects either it is in house development environment or virtual team both will follow different approaches.

## AUTHORS' BACKGROUND

| Your Name | Position | Email | Research Field | Personal website |
|---|---|---|---|---|
| Attique Ur Rehman | Master student | aurehman.cse19ceme@ce.ceme.edu.pk | Software Engineering, Software project management | |
| Ali Nawaz | Master student | anawaz.cse19ceme@ce.ceme.edu.pk | Software Engineering, Machine Learning, Data Mining | |
| Muhammad Abbas | Associate Professor | m.abbas@ceme.nust.edu.pk | ERP Systems, Project Management | |